\documentclass[pre,twocolumn]{revtex4}
\usepackage{graphicx,amssymb}

\begin{document}
\title{Is the entropy $S_q$ extensive or nonextensive? \footnote{To appear in the Proceedings of the 31st Workshop of the International School of Solid State Physics ``Complexity, Metastability and Nonextensivity", held at the Ettore Majorana Foundation and Centre for Scientific Culture, Erice (Sicily) in 20-26 July 2004, eds. C. Beck, A. Rapisarda and C. Tsallis (World Scientific, Singapore, 2005).}
}

\author{
Constantino Tsallis \footnote{tsallis@santafe.edu, tsallis@cbpf.br} 
}

\address{Santa Fe Institute,
1399 Hyde Park Road,
Santa Fe, New Mexico 87501,  USA \\
and\\
Centro Brasileiro de Pesquisas Fisicas, Rua Xavier Sigaud 150, 
22290-180 Rio de Janeiro-RJ, Brazil
}
\date{\today}
\begin{abstract}
The cornerstones of Boltzmann-Gibbs and nonextensive statistical mechanics respectively are the  entropies   $S_{BG} \equiv\ -k \sum_{i=1}^W p_i \ln p_i $ and $S_{q}\equiv k\,(1-\sum_{i=1}^Wp_i^{\;q})/(q-1)\;\;(q\in{\mathbb R}\,;\;S_1=S_{BG})$.   Through them we revisit the concept of additivity,   and illustrate the (not always clearly perceived) fact that (thermodynamical) extensivity has a well defined sense {\it only} if we specify the composition law that is being assumed for the subsystems (say  $A$ and $B$). If the composition law is {\it not} explicitly indicated, it is {\it tacitly} assumed that $A$ and $B$ are {\it statistically independent}. In this case, it immediately follows that $S_{BG}(A+B)= S_{BG}(A)+S_{BG}(B)$, hence extensive, whereas $S_q(A+B)/k=[S_q(A)/k]+[S_q(B)/k]+(1-q)[S_q(A)/k][S_q(B)/k]$, hence nonextensive for $q \ne 1$. In the present paper we illustrate the remarkable changes that occur when $A$ and $B$ are {\it specially correlated}. Indeed, we show that, in such case, $S_q(A+B)=S_q(A)+S_q(B)$ for the appropriate value of $q$ (hence extensive), whereas $S_{BG}(A+B)  \ne S_{BG}(A)+S_{BG}(B)$ (hence nonextensive).
We believe that these facts substantially improve the understanding of the mathematical need and physical origin of nonextensive statistical mechanics, and its interpretation in terms of effective occupation of the $W$ {\it a priori} available microstates of the full phase space. In particular, we can appreciate the origin of the following important fact. In order to have entropic extensivity (i.e., $\lim_{N\to\infty}S(N)/N<\infty$, where $N\equiv number\;of\; elements\; of\; the \;system$), we must use (i)  $S_{BG}$, if the number $W^{\mbox{\it eff}}$ of {\it effectively occupied} microstates increases with $N$ like $W^{\mbox{\it eff}} \sim W \sim \mu^N\;\;(\mu \ge 1)$; (ii) $S_q$ with $q=1-1/\rho$,
 if $     W^{\mbox{\it eff}}\sim N^\rho<W \;\;(\rho \ge 0)$. We had previously conjectured the existence of these two markedly different classes. The contribution of the present paper is to illustrate, for the first time as far as we can tell, the derivation of these facts {\it directly} from the set of probabilities of the $W$ microstates. 

\end{abstract}
\maketitle

\section{Introduction}
\label{section_introduction}

A quantity $X(A)$ associated with a system $A$ is said {\it additive} with regard to a (specific) composition of $A$ and $B$ if it satisfies 
\begin{equation}
X(A+B)=X(A)+X(B) \;,
\end{equation}
where $+$ inside the argument of $X$ precisely indicates that composition. For example, suppose we partition the interior of a single closed bottle in two parts. If no chemical or other reactions occur between the gas molecules that might be inside the bottle, nor between these molecules and the bottle itself (and its internal physical partition), the number of gas molecules is an additive quantity with regard to the elimination of the partition surface. The same happens with the total  energy of an ideal gas, where all interactions have been neglected, including the gravitational one. More trivially, the total height of various (rectangular) doors is, practically speaking, an additive quantity, {\it if} we pile them one above the other one. Not so if we put them side by side! On an abstract level, it is clear that this additivity just corresponds to the number of elements of the union of two sets $A$ and $B$ that have no common elements. 

If, instead of two subsystems $A$ and $B$, we have $N$ of them ($A_1, A_2, ..., A_N$), then we have that
\begin{equation}
X(\sum_{i=1}^N A_i)=\sum_{i=1}^N X(A_i) \;.
\end{equation}
If the subsystems happen to be all equal (a quite common case), then we have that
\begin{equation}
X(N)=NX(1)\;,
\end{equation}
with the notations $X(N) \equiv X(\sum_{i=1}^N A_i)$ and $X(1) \equiv X(A_1)$. 

An intimately related concept is that of {\it extensivity}. It appears frequently in thermodynamics and elsewhere, and corresponds to a weaker demand, namely that of
\begin{equation}
\lim_{N \to\infty}\frac{|X(N)|}{N} < \infty \,.
\end{equation}
Clearly, all quantities that are additive with regard to a given composition, also are extensive with regard to that same composition (and $\lim_{N \to\infty}X(N)/N=X(1)$), whereas the opposite is not necessarily true. For example, the total energy, the total entropy and the total magnetization of the standard Ising ferromagnetic model with $N$ spins on a square lattice are extensive but not additive quantities. In other words, they are asymptotically additive, but not strictly additive. Of course, there are quantities that are neither additive nor even extensive. They are called {\it nonextensive}. All types of behaviors can exist, such as $X(N) \propto N^\gamma \;\;(\gamma \ge 0)$. For instance, thermodynamical quantities that, with regard to some specific composition, exhibit $\gamma=0$ are called {\it intensive}. Such is the case of the temperature, pressure, chemical potential and similar quantities in a great variety of (thermodynamically equilibrated) systems observed in nature. A less trivial example of nonextensive quantity emerges within a spatially homogeneous $d-$dimensional classical gas whose $N$ particles (exclusively) interact through a two-body interaction potential that is strongly repulsive at short distances whereas it is attractive at long distances, decaying like $1/r^\alpha$ ($r \equiv distance \,between\, two\, particles$), and $0 \le \alpha/d $. The total potential energy of such a system corresponds \cite{tsallisvelho} to $\gamma = 2-\alpha/d$ if $0 \le \alpha/d<1$ (i.e., nonextensive), and to $\gamma=1$ for $\alpha/d >1$ (i.e., extensive). The total potential energy of this particular model has a logarithmic $N$-dependance (i.e., nonextensive) at the limiting value $\alpha/d=1$. The Lennard-Jones model for gases corresponds to $(\alpha,d)=(6,3)$, and has therefore an extensive total energy. In contrast, if we assume a cluster of stars gravitationally interacting (together with some physical mechanism effectively generating repulsion at short distances), we have $(\alpha,d)=(1,3)$, hence nonextensivity for the total potential energy. The physical nonextensivity which naturally emerges in such anomalous systems is, in some theoretical approaches, desguised by artificially dividing the two-body coupling constant (which has in fact no means of ``knowing" the total number of particles of the entire system) by $N^{1-\alpha/d}$. For the particular case $\alpha=0$ this yields the widely (and wildly!) used division by $N$ of the coupling constant, typical for a variety of mean field approaches. See \cite{AnteneodoTsallis} for more details. 

Boltzmann-Gibbs ($BG$) statistical mechanics is based on the entropy
\begin{equation}
S_{BG} \equiv\ -k \sum_{i=1}^W p_i \ln p_i \;,
\end{equation}
with
\begin{equation}
\sum_{i=1}^W p_i=1 \;,
\end{equation}
where $p_i$ is the probability associated with the $i^{th}$ microscopic state of the system, and $k$ is Boltzmann constant. In the particular case of equiprobability, i.e., $p_i=1/W$  $(\forall i)$, Eq. (5) yields the celebrated {\it Boltzmann principle} (as named by Einstein \cite{einstein}):
\begin{equation}
S_{BG}=k \ln W \;.
\end{equation}
From now on, and without loss of generality, we shall take $k$ equal to unity.  

Nonextensive statistical mechanics, first  introduced in 1988 \cite{Tsallis88,CuradoTsallis91,TsallisMendesPlastino98} (see \cite{SalinasTsallis,AbeOkamoto,KaniadakisLissiaRapisarda,GrigoliniTsallisWest,Sugiyama,GellMannTsallis,SwinneyTsallis,Kaniadakis,CuradoHerrmannBarbosa} for reviews),  is based on the so-called ``nonextensive" entropy $S_q$ defined as follows:
\begin{equation}
S_{q}\equiv\frac{1-\sum_{i=1}^Wp_i^{\;q}}{q-1}\;\;\;
(q\in{\mathbb R};\;S_1=S_{BG}) \;.
\label{q_entropy}
\end{equation}

For equiprobability (i.e., $p_i=1/W,\,\forall i$), Eq. (8) yields
\begin{equation}
S_q=\ln_q W \;,
\end{equation}
with the {\it $q$-logarithm} function defined \cite{quimicanova} as
\begin{equation}
\ln_q z \equiv \frac{z^{1-q}-1}{1-q} \;\;\;(z \in{\mathbb R}; \;z>0; \;\ln_1 z=\ln z) \;.
\end{equation}
The inverse function, the {\it $q$-exponential}, is given by
\begin{equation}
e_q^z \equiv [1+(1-q)z]^{1/(1-q)} \;\;\;(e_1^z=e^z) 
\end{equation}
if the argument $1+(1-q)z$ is positive, and equals zero otherwise.

The present paper is entirely dedicated to the analysis of the {\it additivity} or {\it nonadditivity} of $S_{BG}$ and of its generalization $S_q$. However, following a common (and sometimes dangerous) practice, we shall from now on cease distinguishing between {\it additive} and {\it extensive}, and use exclusively the word {\it extensive} in the sense of {\it strictly} additive. 

\section{The case of two subsystems}

Consider two systems $A$ and $B$ having respectively $W_A$ and $W_B$ possible microstates. The total number of possible microstates for the system $A+B$ is then {\it in principle} $W \equiv W_{A+B}=W_AW_B$. We emphasized the expression ``in principle" because, as we shall see, a more or less severe reduction of the full phase space might occur in the presence of strong correlations between $A$ and $B$. 

We shall use the notation $p_{ij}^{A+B} \;\;(i=1,2,...,W_A;\; j=1,2,...,W_B)$ for the {\it joint probabilities}, hence
\begin{equation}
\sum_{i=1}^{W_A}\sum_{j=1}^{W_B} p_{ij}^{A+B} =1 \;.
\end{equation}
The {\it marginal probabilities} are defined as follows:
\begin{equation}
p_i^A \equiv \sum_{j=1}^{W_B} p_{ij}^{A+B} \,,
\end{equation}
hence
\begin{equation}
\sum_{i=1}^{W_A} p_i^A =1 \,,
\end{equation}
and
\begin{equation}
p_i^B \equiv \sum_{i=1}^{W_A} p_{ij}^{A+B} \;,
\end{equation}
hence
\begin{equation}
\sum_{j=1}^{W_B} p_j^B =1 \,.
\end{equation} 
These quantities are indicated in the following Table.

\begin{center}
\begin{tabular}{c||c|c|c|c||c}
 $_A\setminus^B$    &  1                          & 2                            &$\;\;\;\;...\;\;\;\;$         &$W_B$                                   \\[1mm] \hline\hline
1           &  $\;\;p_{11}^{A+B}\;\;$     & $\;\;p_{12}^{A+B}\;\;$       &...                     & $\;\;p_{1W_B}^{A+B}\;\;$                                             & $\;\;p_1^A\;\;$   \\[3mm] \hline
2           &  $p_{21}^{A+B}$             & $p_{22}^{A+B}$               &...        & $\;\;p_{2W_B}^{A+B}\;\;$                                              & $p_2^A$   \\[3mm] \hline
...          &  $\;\;...\;\;$                        &  $\;\;...\;\;$                        &  $\;\;...\;\;$       &  $\;\;...\;\;$                                                                   &  $\;\;...\;\;$    \\[3mm] \hline   
$W_A$  &$\;\;p_{W_A1}^{A+B}\;\;$  &$\;\;p_{W_A2}^{A+B}\;\;$  &  $\;\;...\;\;$       &$\;\;p_{W_AW_B}^{A+B}\;\;$        &$p_{W_A}^A$                    \\[3mm] \hline \hline
             &  $p_1^B$                         & $p_2^B$                          &...                     &$p_{W_B}^B$                                                                   & 1
\end{tabular}
\end{center}

We shall next  illustrate the importance of the specification of the composition law. Let us consider two cases, namely independent and (specially) correlated subsystems.

\subsection{Two independent subsystems}

Consider a system composed by two independent subsystems $A$ and $B$, i.e., such that the joint probabilities are given by
\begin{equation}
p_{ij}^{A+B}=p_i^A p_j^B \;\;\;(\forall (i,j)) \;.
\end{equation}
With the definitions
\begin{equation}
S_{BG}(A+B)\equiv -\sum_{i=1}^{W_A}\sum_{j=1}^{W_B} p_{ij}^{A+B} \ln p_{ij}^{A+B} \;,
\end{equation}
\begin{equation}
S_{BG}(A)\equiv -\sum_{i=1}^{W_A} p_{i}^{A} \ln p_{i}^{A} \;,
\end{equation}
and
\begin{equation}
S_{BG}(B)\equiv -\sum_{j=1}^{W_B} p_{j}^{B} \ln p_{j}^{B} \;,
\end{equation}
we immediately verify that
\begin{equation}
S_{BG}(A+B)=S_{BG}(A) +S_{BG}(B)
\end{equation}
and, analogously, that
\begin{equation}
S_{q}(A+B)=S_{q}(A) +S_{q}(B)+(1-q)S_q(A)S_q(B) \;.
\end{equation}
Therefore, $S_{BG}$ is extensive. Consistently, $S_q$ is, unless $q=1$, nonextensive. It is in fact from property (22) that the $q \ne 1$ statistical mechanics we are referring to has been named {\it nonextensive}.

\subsection{Two specially correlated subsystems} 

Consider now that $A$ and $B$ are correlated, i.e., 
\begin{equation}
p_{ij}^{A+B} \ne p_i^A p_j^B  \;,
\end{equation}
Assume moreover, for simplicity, that both $A$ and $B$ systems are equal, and that $W_A=W_B=2$. Assume finally that the joint probabilities are given by the following Table (with $1/2<p<1$): 
\begin{center}
\begin{tabular}{c||c|c||c}
 $_A\setminus^B$    &  1                          & 2                                                    \\[1mm] \hline\hline
1  &  $\;\;2p-1\;\;$                 & $\;\;1-p\;\;$   & $\;\;p\;\;$   \\[3mm] \hline
2  &  $1-p$                           & 0                  & $1-p$   \\[3mm] \hline \hline
    &  $p$                              & $1-p$           & 1
\end{tabular}
\end{center}

It can be trivially verified that Eq. (21) is not satisfied. Therefore, for this special correlation, $S_{BG}$ is {\it  nonextensive}. It can also be verified that, for $q=0$ and only for $q=0$, the following additivity is satisfied:
\begin{equation}
S_0(A+B)=S_0(A) + S_0(B) \,,
\end{equation}
therefore $S_0$ is {\it extensive}. Indeed $S_0(A+B)=2S_0(A)=2$. We immediately see that, depending on the type of correlation (or lack of it) between $A$ and $B$, the entropy which is extensive (reminder: as previously announced, we are using here and in the rest of the paper ``extensive" to strictly mean ``additive") {\it can} be $S_{BG}$ or a different one.

Before going on, let us introduce right away the distinction between {\it a priori possible} states (in number $W$) and {\it allowed} or {\it effective} states (in number $W^{\mbox{\it eff}}$). Let us consider the above case of two equal binary subsystems $A$ and $B$ and consequently  $W=4$. If they are independent (i.e., the $q=1$ case), their generic case corresponds to $0<p<1$, hence  $W^{\mbox{\it eff}}=4$. But if they have the above special correlation (i.e., the $q=0$ case), their generic case corresponds to $1/2<p<1$, hence $W^{\mbox{\it eff}}=3$ (indeed, the state (2,2), although possible a priori, {\it has zero probability}). This type of distinction is at the basis of this entire paper. Notice also that the $q=1$ and $q=0$ cases can be unified through $W^{\mbox{\it eff}}=[2^{1-q}+2^{1-q}-1]^{1/(1-q)}=[2^{2-q}-1]^{1/(1-q)}$. This specific unification will be commented later on.

Let us further construct on the above observations. Is it possible to unify, at the level of the joint probabilities, the case of independence (which corresponds to $q=1$) with the specially correlated case that we just analyzed (which corresponds to $q=0$)? {\it Yes, it is possible}. Consider the following Table:
\begin{center}
\begin{tabular}{c||c|c||c}
 $_A\setminus^B$    &  1                          & 2                                                    \\[1mm] \hline\hline
1  &  $\;\;f_q(p)\;\;$              & $\;\;p-f_q(p)\;\;$            & $\;\;p\;\;$   \\[3mm] \hline
2  &  $p-f_q(p)$                   & $1-2p+f_q(p)$              & $1-p$   \\[3mm] \hline \hline
   &  $p$                               & $1-p$                           & 1
\end{tabular}
\end{center}
where $f_q(p)$ is given by the following relation:
\begin{equation}
2p^q+2(1-p)^q-(f_q)^q-2(p-f_q)^q-(1-2p+f_q)^q=1\,,
\end{equation}   
with $f_q(1)=1$, and $0 \le q \le 1$ (later on we shall comment on  values outside this interval). Typical curves $f_q(p)$ are indicated in Fig. 1. Since Eq. (25) is an implicit one, they have been calculated numerically. It can be checked, for instance, that $f_q(1/2)$ smoothly increases from zero to 1/4 when $q$ increases from zero to unity, being very flat in the neighborhood of $q=0$, and rather steep in the neighborhood of $q=1$. The interesting point, however, is that it can be straightforwardly verified that, for the value of $q$ chosen in $f_q(p)$ defined through Eq. (25) (and only for that $q$),
\begin{equation}
S_q(A+B)=2S_q(A)= 2\frac{1-p^q-(1-p)^q}{q-1} \;,
\end{equation}   
where we have used the fact that $A=B$. In other words, we are facing a whole family of entropies that are {\it extensive} for the respective special correlations indicated in the  Table just above.

\begin{figure}
\begin{center}
\includegraphics[width=\columnwidth,angle=0]{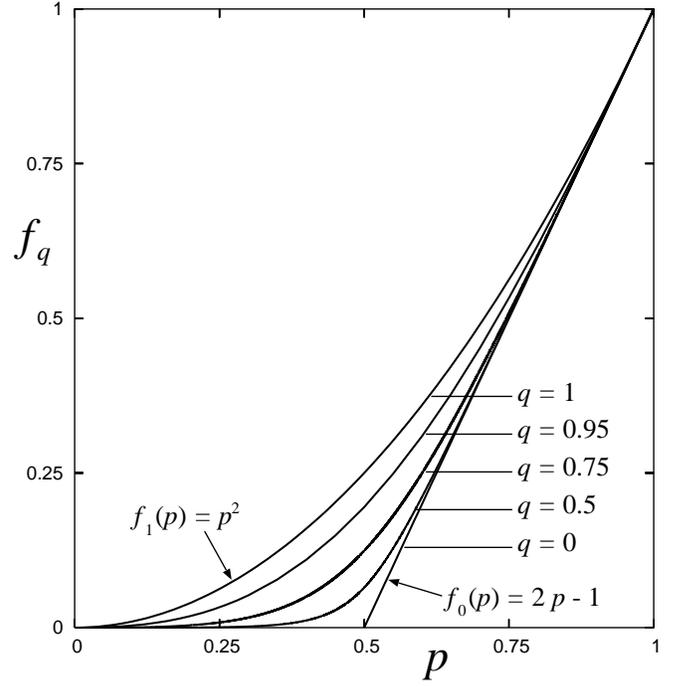}
\end{center}
\caption{\small 
The function $f_q(p)$, corresponding to the two-system $A=B$ case (with $W_A=W_B=2$),  for typical values of $q \in [0,1]$. A few typical nontrivial $(q,f_q(1/2))$ points are $(0.4,0.043295),(0.5,0.064765),(0.6,0.087262),(0.7,0.111289)$, $(0.8,0.138255),(0.9,0.171838),(0.99,0.225630)$. It can be easily verified that these values satisfy the relation $2^{1-q}-[f_q(1/2)]^q-[(1/2) - f_q(1/2)]^q=1/2$, which is the simple form that takes Eq. (25) for the $p=1/2$ particular case. We also remind the trivial values $f_0(1/2)=0$ and $f_1(1/2)=1/4$.
}
\end{figure} 

Let us proceed and generalize the previous examples to two-state systems $A$ and $B$ that are not necessarily equal. The case of independence is trivial, and is indicated in the following Table:
\begin{center}
\begin{tabular}{c||c|c||c}
 $_A\setminus^B$    &  1                          & 2                                                    \\[1mm] \hline\hline
1  &  $\;\;p_1^Ap_1^B\;\;$   & $\;\;p_1^Ap_2^B\;\;$   & $\;\;p_1^A\;\;$   \\[3mm] \hline
2  &  $p_2^Ap_1^B$           & $p_2^Ap_2^B$           & $p_2^A$   \\[3mm] \hline \hline
   &  $p_1^B$                       & $p_2^B$                       & 1
\end{tabular}
\end{center}
Of course, Eq. (21) is satisfied. 

Let us consider now the following Table (with $p_1^A+p_1^B>1$):
\begin{center}
\begin{tabular}{c||c|c||c}
 $_A\setminus^B$    &  1                          & 2                                                    \\[1mm] \hline\hline
1  &  $\;\;p_1^A+p_1^B-1\;\;$   & $\;\;1-p_1^B\;\;$   & $\;\;p_1^A\;\;$   \\[3mm] \hline
2  &  $1-p_1^A$                       & 0                           & $1-p_1^A$   \\[3mm] \hline \hline
   &  $p_1^B$                           & $1-p_1^B$                    & 1
\end{tabular}
\end{center}
We verify that Eq. (24) is satisfied. Is it possible to unify the above anisotropic $q=1$ and $q=0$ cases? {\it Yes, it is}. The special correlations for these cases are indicated in the following Table:
\begin{center}
\begin{tabular}{c||c|c||c}
 $_A\setminus^B$    &  1                          & 2                                                    \\[1mm] \hline\hline
1  &  $\;\;f_q(p_1^A,p_1^B)\;\;$              & $\;\;p_1^A-f_q(p_1^A,p_1^B)\;\;$            & $\;\;p_1^A\;\;$   \\[3mm] \hline
2  &  $p_1^B-f_q(p_1^A,p_1^B)$                   & $1-p_1^A-p_1^B+f_q(p_1^A,p_1^B)$              & $1-p_1^A$   \\[3mm] \hline \hline
   &  $p_1^B$                               & $1-p_1^B$                           & 1
\end{tabular}
\end{center}
where $f_q(p_1^A,p_1^B)=  f_q(p_1^B,p_1^A)$, $f_q(p,1)=p$, $f_q(p,p)=f_q(p)$, $f_1(p_1^A,p_1^B)=p_1^Ap_1^B$, and $f_0(p_1^A,p_1^B)=p_1^A+p_1^B-1$. For any value of $q$ in the interval $[0,1]$, and for any probabilistic pair $(p_1^A,p_1^B)$, the function $f_q(p_1^A,p_1^B)$ is (implicitly) defined through
\begin{eqnarray}
(p_1^A)^q+(1-p_1^A)^q+(p_1^B)^q+(1-p_1^B)^q    \nonumber \\
-[f_q(p_1^A,p_1^B)]^q     \nonumber \\
-[p_1^A-f_q(p_1^A,p_1^B)]^q  
-[p_1^B-f_q(p_1^A,p_1^B)]^q   \nonumber \\
-[1-p_1^A-p_1^B + f_q(p_1^A,p_1^B)]^q=1
\end{eqnarray}
(We remind that, for the $q=0$ particular case, it must be $p_1^A+p_1^B>1$). We notice that the special correlations we are addressing here make that {\it all joint probabilities can be expressed as functions of only one of them}, say $p_{11}^{A+B}$, which is determined once for ever. More explicitly, we have that $p_{12}^{A+B}=p_1^A - p_{11}^{A+B},\, p_{21}^{A+B}= p_1^B - p_{11}^{A+B},\, p_{22}^{A+B}= 1-p_1^A-p_1^B-p_{11}^{A+B} $. 

Eq. (27) recovers Eq. (25) as the particular instance $p_1^A=p_1^B$. And we can easily verify that, for $0 \le q \le1$, 
\begin{equation}
S_q(A+B)=S_q(A)+S_q(B) \;.
\end{equation}
So, we still have extensivity for the appropriate value of $q$, i.e., the value of $q$ which has been chosen in Eq. (27) to define the function $f_q(x,y)$ reflecting the special type of correlations assumed to exist between $A$ and $B$. In other words, when the marginal probabilities have all the information, then the appropriate entropy is $S_{BG}$. But this happens {\it only} when $A$ and $B$ are independent. In all the other cases addressed within the above Table, the important information is by no means contained in the marginal probabilities, and we have to rely on the full set of joint probabilities. In such cases, $S_{BG}$ is {\it nonextensive}, whereas $S_q$ is {\it extensive}. 

Before closing this section dedicated to the case of two systems, let us indicate the Table associated to the $q=0$ entropy for arbitrary systems $A$ and $B$:
\begin{center}
\begin{tabular}{c||c|c|c|c||c}
 $_A\setminus^B$    &  1                          & 2                            &$\;\;\;\;...\;\;\;\;$         &$W_B$                                   \\[1mm] \hline\hline
1           &  $\;\;p_1^A  +p_1^B-1\;\;$     & $\;\;p_2^B\;\;$             &...                           & $\;\;p_{W_B}^B\;\;$                                             & $\;\;p_1^A\;\;$   \\[3mm] \hline
2           &  $p_2^A$                              & 0           &...                        & 0                                              & $p_2^A$   \\[3mm] \hline
...          &  $\;\;...\;\;$                            &  $\;\;...\;\;$                        &  $\;\;...\;\;$       &  $\;\;...\;\;$                                                                   &  $\;\;...\;\;$    \\[3mm] \hline   
$W_A$  &$\;\;p_{W_A}^A\;\;$              &0  &  $\;\;...\;\;$       &0        &$p_{W_A}^A$                    \\[3mm] \hline \hline
             &  $p_1^B$                             & $p_2^B$                          &...                     &$p_{W_B}^B$                                                                   & 1
\end{tabular}
\end{center}
We easily verify that Eq. (24) is satisfied. For example, the generic case corresponds to all probabilities in the Table being nonzero, excepting those explicitly indicated in the Table. For this case we have $S_0(A)=W_A-1$, $S_0(B)=W_B-1$, and $S_0(A+B)=W_A+W_B -2$. This is a neat illustration of the fact that, although the full space admits in principle $W=W_AW_B$ microstates, the strong correlations reflected in the Table make that the system uses appreciably less, namely, in this example, $W^{\mbox{\it eff}}= W_A+W_B -1$. It is tempting to conjecture the generalization of  this expression into $W^{\mbox{\it eff}}= [W_A^{1-q}+W_B^{1-q} -1]^{1/(1-q)}$ for $0 \le q \le 1$. It is clear that $W^{\mbox{\it eff}} \le W_AW_B$, the equality holding only for $q=1$. Since, strictly speaking, $W_A$, $W_B$ and $W^{\mbox{\it eff}}$ are integer numbers, this expression for $W^{\mbox{\it eff}}$ can only be generically valid for real $q \ne 0,1$ in some appropriate asymptotic sense. This sense has to be for $W_A,W_B >>1$, which however are not fully addressed in the present paper for $q \ne 0,1$. For the particular instance $A=B$, we have  $W^{\mbox{\it eff}}= [2 W_A^{1-q} -1]^{1/(1-q)}$.

We also verify another interesting aspect. If $A$ and $B$ are independent, equal values in the marginal probabilities are perfectly compatible with equal values in the joint probabilities. In the most general independent two-system case, we can simultaneously have $p_i^A=1/W_A\;(\forall i)$, $p_j^B=1/W_B\;(\forall j)$, and $p_{ij}^{A+B}=1/(W_AW_B)\;(\forall (i,j))$. This is {\it not} possible in the above Table. Indeed, equal probability values for all {\it allowed} microstates in the Table imply $p_{ij}^{A+B}=1/(W_A+W_B-1) \;(\forall (i,j))$, which is incompatible with equal values for the marginal probabilities. This fact starts pointing into what kind of (irreducibly correlated) situation, the usual BG microcanonical hypothesis ``equal probability occupation of the entire phase space" for thermal equilibrium might become inadequate. It is very plausible that a variety of microscopic dynamical situations must exist (e.g., long-range-interacting Hamiltonian systems) for which the standard equilibrium hypothesis is an oversimplification for physically relevant stationary states that do {\it not} correspond to thermal equilibrium.

\section{The case of three subsystems}

Consider now three systems $A$, $B$ and $C$, having respectively $W_A$, $W_B$ and $W_C$ possible microstates. The total number of possible microstates for the system $A+B+C$ is then {\it in principle} $W \equiv W_{A+B+C}=W_AW_BW_C$. As for the case of two systems, we shall see that strong collective correlations between $A$, $B$ and $C$ may cause a severe reduction of the allowed phase space. 

We shall use the notation $p_{ijk}^{A+B+C} \;\;(i=1,2,...,W_A;\; j=1,2,...,W_B; \;k=1,2,...,W_C)$ for the {\it joint probabilities}, hence
\begin{equation}
\sum_{i=1}^{W_A}\sum_{j=1}^{W_B} \sum_{k=1}^{W_C}p_{ijk}^{A+B+C} =1 \;.
\end{equation}
The $AB-${\it marginal probabilities} are defined as follows:
\begin{equation}
p_{ij}^{A+B} \equiv \sum_{k=1}^{W_C}        p_{ijk}^{A+B+C} \,,
\end{equation}
hence
\begin{equation}
\sum_{i=1}^{W_A}      \sum_{j=1}^{W_B} p_{ij}^{A+B} =1 \,.
\end{equation}
Similar expressions exist for the $AC-$ and $BC-${\it marginal probabilities}. 
The joint probabilities for the $W_A=W_B=W_C=2$ case are indicated in the following Table, where the numbers without parentheses correspond to system $C$ in state 1, and the numbers within parentheses correspond to system $C$ in state 2.

\begin{center}
\begin{tabular}{c||c|c||}
 $_A\setminus^B$    &  1                                    & 2                                                    \\
[1mm] \hline\hline
1                              &  $\;\;p_{111}^{A+B+C}\;\;$                           & $\;\;p_{121}^{A+B+C}\;\;$    \\   
                                &$(p_{112}^{A+B+C})$                                  &$(p_{122}^{A+B+C})$            \\                                               
[3mm] \hline
2                              &  $p_{211}^{A+B+C}$                                    & $p_{221}^{A+B+C}$             \\
                                &  $(p_{212}^{A+B+C})$                                  & $(p_{222}^{A+B+C})$          \\                                         
[3mm] \hline \hline
\end{tabular}
\end{center}

The corresponding $AB-$marginal probabilities are indicated in the Table below:
\begin{center}
\begin{tabular}{c||c|c||}
 $_A\setminus^B$    &  1                                    & 2                                                    \\
[1mm] \hline\hline
1                              &  $\;\;p_{11}^{A+B}\;\;$                           & $\;\;p_{12}^{A+B}\;\;$    \\                                                
[3mm] \hline
2                              &  $p_{21}^{A+B}$                                    & $p_{22}^{A+B}$             \\                                       
[3mm] \hline \hline
\end{tabular}
\end{center}
which of course reproduces the situation we had for the two-system ($A+B$) problem. This is to say $p_{11}^{A+B}=p_{111}^{A+B+C} + p_{112}^{A+B+C}$, and so on.

\subsection{Three independent subsystems}

Consider first the case where all three subsystems $A$ and $B$ are binary and statistically independent, i.e., such that the joint probabilities are given by
\begin{equation}
p_{ijk}^{A+B+C}=p_i^A p_j^Bp_k^C \;\;\;(\forall (i,j,k)) \;.
\end{equation}
The corresponding Table is of course as follows
\begin{center}
\begin{tabular}{c||c|c||}
 $_A\setminus^B$    &  1                                                                   & 2                                                    \\
[1mm] \hline\hline
1                              &  $\;\;p_1^Ap_1^Bp_1^C\;\;$                            & $\;\;p_1^Ap_2^Bp_1^C\;\;$    \\   
                                & $(p_1^Ap_1^Bp_2^C)$                                  &$(p_1^Ap_2^Bp_2^C)$            \\                                               
[3mm] \hline
2                              &  $p_2^Ap_1^Bp_1^C$                                    & $p_2^Ap_2^Bp_1^C$             \\
                                &  $(p_2^Ap_1^Bp_2^C)$                                  & $(p_2^Ap_2^Bp_2^C)$          \\                                         
[3mm] \hline \hline
\end{tabular}
\end{center}
We immediately verify that
\begin{equation}
S_{BG}(A+B+C)=S_{BG}(A) +S_{BG}(B)+S_{BG}(C)
\end{equation}
Therefore, $S_{BG}$ is extensive. Consistently, $S_q$ is, unless $q=1$, nonextensive. 

\subsection{Three specially correlated subsystems} 

Consider now that the three binary subsystems are correlated as indicated in the next Table (with $p_1^A + p_1^B + p_1^C > 2$): 
\begin{center}
\begin{tabular}{c||c|c||}
 $_A\setminus^B$    &  1                                    & 2                                                    \\
[1mm] \hline\hline
1                              &  $\;\;p_1^A+p_1^B+p_1^C-2\;\;$                           & $\;\;1-p_1^B\;\;$    \\   
                                & $(1-p_1^C)$                                                        &$(0)$            \\                                               
[3mm] \hline
2                              &  $1-p_1^A$                                                         & $0$             \\
                                &  $(0)$                                  & $(0)$          \\                                         
[3mm] \hline \hline
\end{tabular}
\end{center}
We easily verify that
\begin{equation}
S_{0}(A+B+C)=S_{0}(A) +S_{0}(B)+S_{0}(C) \,.
\end{equation}
For example, if $A=B=C$ and $2/3<p<1$, we have that $S_0(A+B+C)=3 S_0(A)=3$.

Let us next unify the $q=1$ and the $q=0$ cases. We heuristically found the solution. It is indicated in the following Table:
\begin{center}
\begin{tabular}{c||c|c||}
 $_A\setminus^B$    &  1                                                                                        & 2                                             \\
[1mm] \hline\hline
1                              &  $f_q(p_1^A,p_1^C)+f_q(p_1^B,p_1^C)$                            & $-f_q(p_1^A,p_1^B)$                    \\   
                                &  $ -p_1^C(p_1^A+p_1^B)$                                                  & $   +p_1^A(p_1^B+p_1^C)$             \\
                                &  $  +p_1^Cf_q(p_1^A,p_1^B)  $                                         & $ -p_1^Af_q(p_1^B,p_1^C)$                                \\
                                &                                                                                             &                                               \\
                                &  $[f_q(p_1^A,p_1^B)+p_1^C(p_1^A+p_1^B)$                      &$[p_1^A(1-p_1^B-p_1^C$            \\      
                                &  $-f_q(p_1^A,p_1^C)-f_q(p_1^B,p_1^C)$                             & $+f_q(p_1^B,p_1^C))]$            \\
                                &  $-p_1^Cf_q(p_1^A,p_1^B)]$                                                &                                                                    \\
[3mm] \hline
2                              &  $-f_q(p_1^A,p_1^B)+p_1^B(p_1^A+p_1^C)$                       & $p_1^C(1-p_1^A-p_1^B$               \\
                                &  $-p_1^Bf_q(p_1^A,p_1^C)$                                                  & $+f_q(p_1^A,p_1^B))$                 \\
                                &                                                                                             &                                               \\
                                &  $[p_1^B(1-p_1^A-p_1^C$                                                       & $[(1-p_1^C)(1-p_1^A-p_1^B$             \\     
                                &  $+f_q(p_1^A,p_1^C))]$                                                       & $+f_q(p_1^A,p_1^B))]$            \\                        
[3mm] \hline \hline
\end{tabular}
\end{center}
where the function $f_q(x,y)$ is defined in Eq. (27). Interestingly enough, it has been possible to find a three-subsystem solution in terms of the two--subsystem and one-system ones.  More explicitly, we have, for example, that $p_{111}^{A+B+C}= f_q(p_1^A,p_1^C)+f_q(p_1^B,p_1^C)-p_1^C(p_1^A+p_1^B)+p_1^Cf_q(p_1^A,p_1^B)  = p_{11}^{A+C} +p_{11}^{A+B} -p_1^C(p_1^A+p_1^B) +p_1^C p_{11}^{A+B} $, and similarly for the other seven three-subsystem joint probabilities. Of course, all eight joint probabilities associated with the above Table are nonnegative; whenever the  values of $(p_1^A,p_1^B,p_1^C)$ replaced within one or the other of these analytic expressions yield negative numbers, the corresponding probabilities are to be taken equal to zero. 

The $AB-$marginal probabilities precisely recover the joint probabilities of the previously discussed two-system ($A+B$) Table. For example, $[f_q(p_1^A,p_1^C)+f_q(p_1^B,p_1^C)-p_1^C(p_1^A+p_1^B)+p_1^Cf_q(p_1^A,p_1^B)] + [f_q(p_1^A,p_1^B)+p_1^C(p_1^A+p_1^B)-f_q(p_1^A,p_1^C)-f_q(p_1^B,p_1^C)-p_1^Cf_q(p_1^A,p_1^B)] = f_q(p_i^A,p_1^B)$, $[-f_q(p_1^A,p_1^B)+p_1^A(p_1^B+p_1^C)-p_1^Af_q(p_1^B,p_1^C)] + [p_1^A(1-p_1^B-p_1^C+f_q(p_1^B,p_1^C))] = p_1^A - f_q(p_1^A,p_1^B)$, and so on.

Finally, we  verify that
\begin{eqnarray}
S_q(A+B+C) &=& \frac{1}{2} [S_q(A+B)+S_q(A+C) + S_q(B+C)] \nonumber \\
&=&  S_q(A) +S_q(B)+S_q(C)
\end{eqnarray}

For the particular case $A=B=C$, the above Table becomes

\begin{center}
\begin{tabular}{c||c|c||}
 $_A\setminus^B$    &  1                                                                & 2                                                    \\
[1mm] \hline\hline
1                              &  $2(f_q(p)-p^2)+pf_q(p)$                            & $2p^2-f_q(p)-pf_q(p)$    \\   
                                & $[2p^2-f_q(p)-pf_q(p)]$                               &$[p(1-2p+f_q(p))]$            \\                                               
[3mm] \hline
2                              &  $2p^2-f_q(p)-pf_q(p)$                                & $p(1-2p+f_q(p))$             \\
                                &  $[p(1-2p+f_q(p))]$                                      & $[(1-p)(1-2p+f_q(p))]$          \\                                         
[3mm] \hline \hline
\end{tabular}
\end{center}
where we have used $f_q(p,p)=f_q(p)$.

For the generic case of three subsystems with $W_A$, $W_B$ and $W_C$ states respectively, we have that $W=W_A W_B W_C$, whereas in the appropriate asymptotic sense we expect $W^{\mbox{\it eff}}=[W_A^{1-q}+  W_B^{1-q} + W_C^{1-q} -2]^{1/(1-q)} \le W$ for $0 \le q \le 1$ (the equality generically holds only for $q=1$). 
In the particular instance $A=B=C$, this expression becomes $W^{\mbox{\it eff}}=[3 W_A^{1-q} -2]^{1/(1-q)}$.
\section{Enlarging the scenario}

\subsection{The case of $N$ subsystems}

The three-system case discussed above is a generic one under the assumption that $W_A=W_B=W_C=2$. We have not attempted to generalize its corresponding special correlation Table to the generic $(W_A,W_B,W_C)$ case, and even less to the even more generic case of $N$ such systems ($A_1,A_2,...,A_N$). It is clear however that, assuming that this (not necessarily trivial) task was satisfactorily accomplished, the result would lead to
\begin{equation}
S_q(\sum_{r=1}^N A_r)=\sum_{r=1}^NS_q(A_r) \,,
\end{equation}
where $q=1$ {\it if} all $N$ systems are mutually independent, i.e.,
\begin{equation}
p_{i_1i_2...i_N}^{A_1+A_2+...+A_N}=\prod_{r=1}^N p_{i_r}^{A_r} \;\;\;(\forall (i_1,i_2,...,i_N)) \;,
\end{equation}
and $q \ne 1$ otherwise. This is to say, if we have independence, the only entropy which is extensive is $S_{BG}$. If we do {\it not} have independence but the special type of (collective) correlations focused on in this paper instead, then only $S_q$ for a special value of $q$ is extensive. 

For the case of independence, the generic composition law for $S_q$ is given by
\begin{equation}
\ln[1+(1-q)S_q(\sum_{r=1}^N A_r)]=\sum_{r=1}^N \ln[1+(1-q)S_q(A_r)] \;, 
\end{equation}
or, equivalently,
\begin{equation}
1+(1-q)S_q(\sum_{r=1}^N A_r)=\prod_{r=1}^N [1+(1-q)S_q(A_r)] \;.
\end{equation}
Eq. (38) exhibits in fact the well known (monotonic) connection between $S_q$ and the Renyi entropy $S_q^R \equiv \Bigl[\ln \sum_{i=1}^W p_i^q \Bigr]/(1-q)=\Bigl[ \ln[1+(1-q)S_q] \Bigr] /(1-q)$ (we remind that, for independent systems, $S_q^R$ is extensive, $\forall q$).

We have generically $W=\prod_{r=1}^N W_{A_r}$, which corresponds of course to the total number of a priori possibly occupied states (i.e., whose joint probabilities are generically nonzero) for the generic $q=1$ case. In contrast, the generic $q=0$ case has only $W^{\mbox{\it eff}}=(\sum_{r=1}^NW_{A_r}) -(N-1)$ nonzero joint probabilities. These are
$p_{11...1}^{A_1+A_2+...+A_N}=( \sum_{r=1}^N p_1^{A_r}) -(N-1) \ge 0$, 
$p_{i_111...1}=p_{i_1}^{A_1} $ $(i_1=2,3,...,W_{A_1})$, 
$p_{1i_211...1}=p_{i_2}^{A_2} $ $(i_2=2,3,...,W_{A_2})$, 
$p_{111...i_N}=p_{i_N}^{A_N} $ $(i_N=2,3,...,W_{A_N})$.
The generic $q=1$ and $q=0$ cases can, analogously to what has been done before, be unified through
$W^{\mbox{\it eff}}=\Bigl[(\sum_{r=1}^N W_{A_r}^{1-q}) -(N-1) \Bigr]^{1/(1-q)} \le W$ ($0 \le q \le 1$), where the equality holds only for $q=1$. In the particular instance $A_1=A_2=...=A_N \equiv A$, this expression becomes $W^{\mbox{\it eff}}=[N W_A^{1-q} -(N-1)]^{1/(1-q)}$.

Furthermore, for $N$ equal subsystems  (a quite frequent case, as already mentioned), Eq. (36) becomes
\begin{equation}
S_q(N)=NS_q(1) \,,
\end{equation}
where the change of notation is transparent. This is an extremely interesting relation since it already has the shape that accomodates well within standard thermodynamics, {\it even if the entropic index $q$ is not necessarily the usual one, i.e., $q=1$}. It is allowed to think that Clausius would perhaps have been as satisfied with this relation as he surely was with the same relation but with $S_{BG}$! One might also quite safely speculate that if the system is such that its Table of joint probabilities is {\it not exactly} of the type we have discussed here, but {\it close} to it, then we might have, not exactly relation (40) but rather only asymptotically $S_q(N) \propto N$. In other words, as long as the system belongs to what we may refer to as the $q-$universality class, we should expect $\lim_{N \to\infty} S_q(N)/N < \infty$, in total analogy with the usual $BG$ case.

To geometrically interpret Eq. (40), we may consider the case of equal probabilities in the {\it allowed} phase space, i.e., in that part of phase space which is expected to have, not necessarily $W$ microstates, but generically $W^{\mbox{\it eff}}$ microstates (with $W^{\mbox{\it eff}} \le W$). The effective number $W^{\mbox{\it eff}}$ is expected (at least in the $N >>1$ limit) to be precisely   the number of all those states that the special collective correlations allow to visit. So, if we assume equal probabilities in Eq. (40) (i.e., $p_{i_1i_2...i_N}^{A_1+A_2+...+A_N}=1/W^{\mbox{\it eff}}$), we obtain
\begin{equation}
\ln_q W^{\mbox{\it eff}} \equiv \frac{(W^{\mbox{\it eff}})^{1-q} - 1}{1 - q}=NS_q(1) \,,
\end{equation}
or, equivalently
\begin{equation}
W^{\mbox{\it eff}} =e_q^{NS_q(1)}     \equiv [1+(1-q)NS_q(1)]^{1/(1-q)} \,. 
\end{equation}
Two cases are possible for this relation, namely $q=1$ and $q<1$. In the first case, we have the usual result
\begin{equation}
W^{\mbox{\it eff}}=W=\mu^N \,,
\end{equation}
with 
\begin{equation}
\mu \equiv e^{S_{BG}(1)} \ge 1 \,.
\end{equation} 
In the second case, we have an unusual  result, namely
\begin{equation}
W^{\mbox{\it eff}} = [1+NS_q(1)/\rho]^\rho \,,
\end{equation}
with 
\begin{equation}
\rho \equiv 1/(1-q) \ge 0 \,. 
\end{equation}
In the $N \to\infty$ limit, this relation becomes the following one:
\begin{equation}
W^{\mbox{\it eff}} \propto N^\rho \,.
\end{equation}
This (physically quite appealing) possibility was informally advanced by us long ago, and formally in \cite{GellMannTsallis}. It has now been obtained along an appropriate probabilistic path.

\subsection{The $q \to -\infty$ case}

From Eq. (46) we expect the $q \to -\infty$ case to correspond to the limiting situation where $W^{\mbox{\it eff}}$ is constant. To realize this situation, let us first consider the $A=B$ two-system case with the following Table ($W_A=W_B$):
\begin{center}
\begin{tabular}{c||c|c|c|c||c}
 $_A\setminus^B$    &  1                          & 2                            &$\;\;\;\;...\;\;\;\;$         &$W_A$                                   \\[1mm] \hline\hline
1           &  $\;\;p_1\;\;$              & $\;\;0\;\;$                 &0                            & $\;\;0\;\;$              & $\;\;p_1\;\;$   \\[3mm] \hline
2           &  $0$                          & $p_2$                     &0                    & $\;\;0\;\;$                                              & $p_2$   \\[3mm] \hline
...          &  $\;\;0\;\;$                        &  $\;\;0\;\;$                        &  $\;\;...\;\;$       &  $\;\;0\;\;$                                                                   &  $\;\;...\;\;$    \\[3mm] \hline   
$W_A$  &$\;\;0\;\;$                &$\;\;0\;\;$  &  $\;\;0\;\;$       &$\;\;p_{W_A}\;\;$        &$p_{W_A}$                    \\[3mm] \hline \hline
             &  $p_1$                         & $p_2$                          &...                     &$p_{W_A}$                                                                   & 1
\end{tabular}
\end{center}
This Table corresponds to $p_{ij}^{A+B}=p_i\, \delta_{ij}$. Its generalization to $N$ equal systems is trivial: $p_{i_1i_2...i_N}=p_{i_1}$ if all $N$ indices coincide, and zero otherwise. The corresponding entropy therefore asymptotically approaches the relation
\begin{equation}
S_{-\infty}(N)=S_{-\infty}(1) \;\;(\forall N),
\end{equation}
thus corresponding to $\rho = 0$ as anticipated. It appears then that all cases equivalent (through permutations) to the above Table, should yield the same limit $q \to -\infty$.

\subsection{Connection with the Borges-Nivanen-Le Mehaute-Wang $q-$product}

Let us mention at this point an interesting connection that can be established between the present problem and the $q-$product introduced by L. Nivanen, A. Le Mehaute and Q.A. Wang and by E.P. Borges \cite{borges}. It is defined as follows:
\begin{equation}
x \times_q y \equiv (x^{1-q}+y^{1-q}-1)^{1/(1-q)} \;\;(x \times_1 y=xy) \,.
\end{equation}
It has the elegant, extensive-like, property
\begin{equation}
\ln_q(x \times_q y)=\ln_qx+\ln_qy \,, 
\end{equation}
to be compared with the by now quite usual, nonextensive-like, property
\begin{equation}
\ln_q(x y)=\ln_qx+\ln_qy+(1-q)(\ln_qx)(\ln_qy) \,.
\end{equation}
This type of structure was since long (at least since 1999) being informally discussed by A.K. Rajagopal, E.K. Lenzi, S. Abe, myself, and probably others. But only very recently it was beautifully formalized \cite{borges}. It has immediately been followed and considerably extended by Suyari in a relevant set of papers \cite{suyari}.  

Let us now go back to the main topic of the present paper. Consider the following joint probabilities associated with $N$ generic subsystems:
\begin{equation}
p_{i_1i_2...i_N}^{A_1+A_2+...+A_N}=\Bigl[1-N+  \phi_{i_1i_2...i_N}^{(q)} +      \sum_{r=1}^N (p_{i_r}^{A_r})^{q-1}\Bigr]^{1/(q-1)} \,,
\end{equation}
where $\phi_{i_1i_2...i_N}^{(q)}$ is a nontrivial function which ensures that 
\begin{equation}
\sum_{i_1i_2...i_N}p_{i_1i_2...i_N}^{A_1+A_2+...+A_N}=1
\end{equation}

In the limit $q \to 1$, Eq. (52) must recover the independent-systems one, namely
\begin{equation}
p_{i_1i_2...i_N}^{A_1+A_2+...+A_N}=\prod_{r=1}^N p_{i_r}^{A_r} \,,
\end{equation}
which implies $\phi_{i_1i_2...i_N}^{(1)}=0$.

Notice that, excepting for the function $\phi_{i_1i_2...i_N}^{(q)}$, Eq. (52) associates $1/p_{i_1i_2...i_N}^{A_1+A_2+...+A_N}$  with $ {\prod_q}_{r=1}^N (1/p_{i_r}^{A_r})$ with ${\prod_q}_{r=1}^N x_r \equiv [{x_1}^{1-q}+{x_2}^{1-q}+...+{ x_N}^{1-q} - N+1]^{1/(1-q)}$ . 

It follows from Eq. (52) that
\begin{eqnarray}
(p_{i_1i_2...i_N}^{A_1+A_2+...+A_N})^q &=& (1-N   +  \phi_{i_1i_2...i_N}^{(q)} )p_{i_1i_2...i_N}^{A_1+A_2+...+A_N}      \nonumber  \\  
&+&  p_{i_1i_2...i_N}^{A_1+A_2+...+A_N}\sum_{r=1}^N (p_{i_r}^{A_r})^{q-1}
\end{eqnarray}
hence
\begin{eqnarray}
\sum_{i_1i_2...i_N}(p_{i_1i_2...i_N}^{A_1+A_2+...+A_N})^q= (1-N)      \nonumber  \\  
+    \sum_{i_1i_2...i_N}    p_{i_1i_2...i_N}^{A_1+A_2+...+A_N}\sum_{r=1}^N (p_{i_r}^{A_r})^{q-1} \;,
\end{eqnarray}
where we have imposed one more nontrival condition on $\phi_{i_1i_2...i_N}^{(q)}$, namely that
\begin{equation}
\sum_{i_1i_2...i_N}p_{i_1i_2...i_N}^{A_1+A_2+...+A_N}\phi_{i_1i_2...i_N}^{(q)} =0 \,.
\end{equation}
One might naturally have the impression that no function $\phi_{i_1i_2...i_N}^{(q)}$ might exist satisfying simultaneously Eqs. (53) and (57). This is not so however, at least for particular cases, since we have explicitly shown in the present paper solutions of this nontrivial problem.

Using the definition of $S_q$ in the left-hand member of the equality we obtain
\begin{eqnarray}
(1-q)S_q(\sum_{r=1}^N A_r)=       \nonumber  \\  
    \sum_{i_1i_2...i_N}    p_{i_1i_2...i_N}^{A_1+A_2+...+A_N}\sum_{r=1}^N (p_{i_r}^{A_r})^{q-1}  -N\;.
\end{eqnarray}
Let us now introduce in Eq. (58) the definition of marginal probabilities, namely
\begin{equation}
p_{i_r}^{A_r}= \sum_{i_1i_2...i_{r-1}i_{r+1}...i_N} p_{i_1i_2...i_N}^{A_1+A_2+...+A_N} \;.
\end{equation}
We obtain
\begin{eqnarray}
(1-q)S_q(\sum_{r=1}^N A_r)=       
    \sum_{r=1}^N (p_{i_r}^{A_r})^q  -N \;.
\end{eqnarray}
Using once again the definition of $S_q$ on the right-hand member, we finally obtain
\begin{equation}
S_q(\sum_{r=1}^N A_r)=\sum_{r=1}^N S_q(A_r) 
\end{equation}
as desired. 

It should, however, be clear that this remarkable mathematical fact by no means exhausts the problem of the search of explicit Tables of joint probabilities that would lead to extensivity of $S_q$ for nontrivial values of $q$. The constraints imposed by the definition itself of the concept of marginal probabilities are of such complexity that the search of solutions is by no means trivial, at least at our present degree of knowledge. Indeed, one easily appreciates this fact by looking at the explicit solutions indicated in Sections II.B and III.B. 

\section{conclusions}


Let us summarize the obvious conclusion of the present paper: {\it Unless the composition law is specified, the question whether an entropy (or some similar quantity) is or is not extensive has no sense}. Allow us a quick digression. The situation is in fact quite analogous to the quick or slow motion of a body. Ancient Greeks considered the motion to be an absolute property. It was not until Galileo that it was clearly perceived that motion has no sense unless the referential is specified. In Galileo's time, and even now, when no referential is indicated, one tacitly assumes that the referential is the Earth. In total analogy, when no composition law is indicated for analyzing the extensivity of an entropy, one tacitly assumes that the subsystems that we are composing are independent. It is {\it only} --- a big {\it only}!--- in this sense that we can say that $S_{BG}$ is extensive, and that $S_q$ (for $q \ne 1$) is nonextensive.

Once we have established the point above, the next natural question is: Are there classes of collective correlations for which we know which is the specific entropy to be  extensive? (knowing, of course, that  absence of all correlations leads to $S_{BG}$). For this operationally important question, nontrivial illustrations on how the entropic form is dictated by the type of special collective correlations that might (or might not) exist in the system have explicitly presented in Section II.B and III.B. From this discussion, two vast categories of systems are identified (at the most microscopic possible level, i.e., that of the joint probabilities), namely those whose {\it allowed} phase space increases (in size) with $N$ like an {\it exponential} or like a {\it power-law}, corresponding respectively to $q=1$ and to $q<1$. 

However, it should be clear that the present paper is only {\it exploratory} in what concerns this hard task. Indeed, we have {\it not} found the generic answer for $N$ (not necessarily equal) systems, and we have basically concentrated {\it only} on the interval $0 \le q \le 1$. We do not even know without doubt if the answer is {\it unique} (excepting of course for trivial permutations), or if it admits a variety of forms all belonging to the same universality class of nonextensivity (i.e., sharing the same value of the entropic index $q$).  Even worse, we still do {\it not} know {\it what} specifically happens in the structure of the allowed phase space in the (thermodynamically) most important limit $N \to \infty$, or in the frequent limit $W_{A} \to\infty$ (which would provide  a precise geometrical interpretation to a formula such as $W^{\mbox{\it eff}}=[N W_A^{1-q} -(N-1)]^{1/(1-q)}$ for say $0 \le q \le 1$).
It is precisely this structure which is crucial for fully understanding nonextensive statistical mechanics and its related applications in terms on nonlinear dynamical systems. For example, an interesting situation might occur if we compare the distribution which optimizes $S_q(N)$ and {\it then} consider $N >>1$, with the distribution corresponding to having {\it first} considered $N>>1$ in $S_q(N)$ and only then optimizing. We certainly expect the thermodynamic limit and the optimization operation to {\it commute} for a system composed by $N$ independent (or nearly independent) subsystems. 
But the situation seems to  be more subtle if our system was composed by $N$ subsystems correlated in that special, collective manner which demands $q \ne 1$ in order to have entropy extensivity. Such a situation would be consistent with a property which emerges again and again \cite{SalinasTsallis,AbeOkamoto,KaniadakisLissiaRapisarda,GrigoliniTsallisWest,Sugiyama,GellMannTsallis,SwinneyTsallis,Kaniadakis,CuradoHerrmannBarbosa}  for nonextensive systems, namely that the $N \to\infty$ and the $t \to\infty$ limits do not necessarily commute. One more relevant issue concerns {\it what} specific dynamical nature is required for a physical system to ``live", in phase space, within a structure close to one of those that we have presently analyzed. It is our conjecture that this would occur for nonlinear dynamical systems whose Lyapunov spectrum is either zero or close to it, i.e., under circumstances similar to the edge of chaos, where many of the so called complex systems are expected to occur. We leave all these questions as open points needing further progress. 

Let us finally mention the following point. It is by no means trivial to find sets of joint probabilities (associated to relevant statistical correlations) that produce very simple marginal probabilities  (such as $p$ and $1-p$ for binary variables) and which {\it simultaneously} admit the imposition  (as we have done here) of strict additivity of the corresponding entropy. This has been possible for $S_q$. This might be in principle possible as well for other entropic forms. The fact however that, like $S_{BG}$,  $S_q$ simultaneously (i) admits such solutions, (ii) is concave ($\forall q>0$), (iii) is Lesche-stable, and (iv) leads to {\it finite} entropy production per unit time \cite{GellMannTsallis}, constitutes --- we believe --- a strong mathematical basis for being physically meaningful in the thermostatistical sense.  

\section*{Acknowledgments}

It is with pleasure that I acknowledge very fruitful discussions with S. Abe, C. Anteneodo, F. Baldovin, E.P. Borges, J.P. Crutchfield, J.D. Farmer, M. Gell-Mann, H.J. Haubold, L. Moyano, A.K. Rajagopal, Y. Sato and D.R. White. I have also benefited from a question put long ago by M.E. Vares related to the possible difference between $W$ and $W^{\mbox{\it eff}}$.

\end{document}